\documentclass[preprint]{aastex}

\def\dfrac#1#2{{\displaystyle\frac{#1}{#2}}}
\def\cfrac#1#2{\dfrac{\mathstrut #1}{#2}}
\def\pa#1#2{\dfrac{\partial #1}{\partial #2}}

\newcommand{\pd}{{\rm d}}

\newcommand{\dlz}{d_{\rm L}(z)}
\newcommand{\zl}{z_{\rm l}}
\newcommand{\zu}{z_{\rm u}}
\newcommand{\Ml}{M_{\rm l}}
\newcommand{\Mu}{M_{\rm u}}
\newcommand{\fobs}{f_{\rm obs}(M, z)}
\newcommand{\nobs}{N_{\rm obs}}
\newcommand{\chis}{\chi_S(M,z)}

\newcommand{\zmax}[1]{z_{{\rm max},#1}}
\newcommand{\zmin}[1]{z_{{\rm min},#1}}
\newcommand{\vmax}[1]{V_{{\rm max}}(#1)}
\newcommand{\phat}{\hat{\phi}}
\newcommand{\vect}[1]{\mbox{\boldmath $#1$}}

\begin{document}

\title{
Tests of Statistical Methods for Estimating Galaxy Luminosity Function and
Applications to the Hubble Deep Field}

\author{
Tsutomu T. Takeuchi\altaffilmark{1}, Kohji Yoshikawa\altaffilmark{1}, 
and Takako T. Ishii\altaffilmark{2}
}

\vspace{6mm}

\affil{
Department of Astronomy, Faculty of Science, Kyoto University,
\\
Sakyo-ku, Kyoto 606--8502, JAPAN
\\
Electronic mail:
takeuchi, kohji, ishii@kusastro.kyoto-u.ac.jp
}

\altaffiltext{1}{
Research Fellows of the Japan Society for the Promotion of Science
}

\altaffiltext{2}{
Kwasan and Hida Observatories, 
Kyoto University, 
Yamashina-ku, Kyoto 607--8471, JAPAN
}

\begin{abstract}
We studied the statistical methods for the estimation of the 
luminosity function  (LF) of galaxies.
We focused on four nonparametric estimators: $1/V_{\rm max}$ estimator, 
maximum-likelihood estimator of Efstathiou et al.~(1988), Cho{\l}oniewski's
estimator, and improved Lynden-Bell's estimator.
The performance of the $1/V_{\rm max}$ estimator has been recently
questioned, especially for the faint-end estimation of the LF.
We improved these estimators for the studies of the distant Universe, 
and examined their performances for various classes of functional forms 
by Monte Carlo simulations.
We also applied these estimation methods to 
the mock 2dF redshift survey catalog prepared by Cole et al. (1998).
We found that $1/V_{\rm max}$ estimator yields a completely unbiased result
if there is no inhomogeneity, but is not robust against clusters or voids.
This is consistent with the well-known results, and we did not confirm 
the bias trend of $1/V_{\rm max}$ estimator claimed by Willmer (1997)
in the case of homogeneous sample.
We also found that the other three maximum-likelihood type estimators are
quite robust and give consistent results with each other.
In practice we recommend Cho{\l}oniewski's estimator for two reasons:
1. it simultaneously provides the shape and normalization of the LF; 
2. it is the fastest among these four estimators, because of the 
algorithmic simplicity.
Then, we analyzed the photometric redshift data of the Hubble Deep Field 
prepared by Fern\'{a}ndez-Soto et al.~(1999) using the above four methods.
We also derived luminosity density $\rho_{\rm L}$ at $B$- and $I$-band.
Our $B$-band estimation is roughly consistent with that of Sawicki, 
Lin, \& Yee (1997), but a few times lower at $2.0 < z < 3.0$.
The evolution of $\rho_{\rm L}(I)$ is found to be less prominent.
\end{abstract}

\keywords{
galaxies: evolution --- galaxies: luminosity function --- 
galaxies: statistics --- methods: statistical 
}

\section{INTRODUCTION}

The luminosity function of galaxies (LF) plays a crucial role for 
extragalactic astronomy and observational cosmology.
It is one of the basic descriptions of the galaxy population
itself, and sometimes treated as a function of color (e.g. Efstathiou, 
Ellis, \& Peterson 1988, hereafter EEP; Metcalfe et al. 1998; 
Lin et al. 1999)
or morphology (e.g. Bingelli, Sandage, \& Tammann 1988; Marzke et al. 
1998), or other additional parameters of galaxies.
It is also essential for interpreting galaxy number counts (e.g. Koo \& 
Kron 1992; Ellis 1997) and for analyzing galaxy clustering (e.g. Strauss \& 
Willick 1995; Efstathiou 1996).
Furthermore, the LF is a fundamental test for the theory of galaxy 
formation (e.g. Baugh, Cole, \& Frenk 1996).
Recently, the exact shape of the LF has been of particular interest, 
because it is one of the key issues to the ``faint blue galaxy problem'' 
of galaxy number counts (Koo \& Kron 1992; Ellis 1997), and may be 
related to dwarf galaxy formation (e.g. Babul \& Rees 1992; Babul \& 
Ferguson 1996; Hogg \& Phinney 1997).
The evolution of the LF is also important to derive the cosmic 
luminosity density, in the context of the `Madau plot', i.e., cosmic star
formation density as a function of redshift 
(e.g. Madau et al. 1996; Cowie et al. 1996; 
Sawicki, Lin, \& Yee 1997, hereafter SLY97; Pascarelle, Lanzetta, \& 
Fern\'{a}ndez-Soto 1998). 

Estimating galaxy luminosity function from an observational galaxy catalog 
is a fundamental work, but it is not a trivial task.
Because of the flux-limited nature of the redshift survey data,
the catalogs are inevitably censored, and suitable statistical technique 
is required.
In the early stage of the extragalactic astronomy, the classical estimator,
the number of galaxies in a given volume, $\Phi = N/V$, was used 
to estimate the LF (Hubble 1936).
Of course this is not sufficient for detailed studies, and many experts
have proposed ingenious methods.
Schmidt (1968) invented the famous $1/V_{\rm max}$ estimator in the studies 
of quasar population.
Felten (1977) introduced the direction dependence of the magnitude limit.
The extension for combining some different catalogs coherently was discussed 
in Avni \& Bahcall (1980).
Further extension to examine the evolution of the LF with redshift was
proposed by Eales (1993), and the integrated variant of the Eales' estimator
was used in the survey of the Hawaii Deep Fields by Cowie et al. (1996).
Qin \& Xie (1999) also developed this estimator with a similar line of 
study.
The fundamental assumption of this estimator is that the distribution of 
the objects is spatially uniform.
Nowadays this is regarded as a drawback, because we know that the galaxies
have strong clustering properties in the large-scale structure.
In spite of the drawback, $1/V_{\rm max}$ estimator has been frequently
used for extragalactic studies (e.g. Lilly et al. 1995; Ellis et al. 1996),
probably because of its simplicity in calculation.

In order to overcome the difficulty in treating inhomogeneous galaxy 
distribution, some density-insensitive methods have been invented.
Lynden-Bell (1971) proposed the $C^-$ method, and applied it to the quasar 
sample of Schmidt (1968).
This method is based on a quite sophisticated statistical idea as we discuss
in subsequent sections.
Carswell (1973) reported numerical experiments in its use.
Jackson (1974) improved the method to combine several different catalogs, 
and studied the error estimation when the LF is expressed as an analytical 
form.
The original method could derive only the shape of the probability density 
function, but Cho{\l}oniewski (1987)(hereafter C87) improved the method to 
obtain the density normalization and to trace the density evolution 
simultaneously.
Lynden-Bell himself, and later Felten (1976) and Nicoll \& Segal (1983) 
pointed out the drawback of this method that it cannot work in the faintest 
regime where the data points are too sparse.
This drawback was basically overcome the introduction of smoothing method 
by Caditz \& Petrosian (1993) (CP93).
Subba Rao et al. (1996) and Szokoly et al.~(1998) used the method in the 
recent studies of distant galaxies.
We note that this method was further generalized by Maloney \& Petrosian (1999)
to treat the doubly truncated data, but here we do not discuss it further.

The method proposed by Turner (1979) and Kirshner, Oemler, \& Schechter 
(1979) used the ratio of the number of objects between the absolute magnitude 
interval $[M, M+\pd M]$ and the number of objects brighter than $M$, 
which canceled out the density inhomogeneity.
Marinoni et al. (1999) used this method in their analysis of the 
effect of the Local infall motion on the estimation of the LF.
Similar estimators were used by Davis \& Huchra (1982) and later 
de Lapparent, Geller, \& Huchra (1989).
However, as mentioned in Efstathiou (1996), this estimator does not use 
the whole sample.

In contrast, some estimation methods using analytical LF models, which are 
often called parametric estimation methods, have been developed.
Sandage, Tammann, \& Yahil (1979)(STY) introduced the maximum 
likelihood method, which was free of the effects induced by density 
inhomogeneity, in this field by using parametric Schechter form for the LF.
This parametric form was extended for evolutionary studies of galaxies 
by Lin et al.~(1999).
Marshall et al. (1983) presented another parametric estimator which 
can treat both the LF and the evolution parameter simultaneously, assuming
the Poisson distribution of the objects on the magnitude--redshift space.

The maximum likelihood approach was widely used and extended to the methods
which did not use analytical forms, often referred to as nonparametric
methods.
Nicoll \& Segal (1983) proposed such a type of estimator and used it for the 
study of their chronometric cosmology.
The estimator which can be regarded as an advanced version of 
Nicoll \& Segal's method
was invented by Cho{\l}oniewski (1986)(C86).
This method adopts the same assumption as Marshall et al.~(1983), and is
regarded as a binned nonparametric version of it.
Another stepwise estimator, which was a binned analog of STY's 
estimator, was introduced by EEP.
Now this method seems to be most commonly used, and is called 
`the stepwise maximum likelihood method'.
But note that not only EEP's but also most of the other 
estimators are based on the maximum likelihood principle.
The EEP's method was extended to treat density evolution 
(Heyl et al.~1997; Springel \& White 1998).

In spite of the variety of the methods, as we see above, there had been
only the comparisons of some methods in the literature (e.g. Felten 1976; 
C86; EEP; Heyl et al. 1997) before the elaborate 
intercomparison by Willmer (1997)(W97).
Statistically detailed discussions are not so frequently seen, either, 
except the rigorous works of Petrosian (1992).
In W97, each method was examined by Monte Carlo simulations and CfA1 (e.g. 
de Lapparent et al. 1989) data.
The obtained results were fitted by Schechter form, and W97 discussed
the distribution of the estimates by each method after 1000 simulations.
Based on the fitting parameter distributions, W97 reported the bias trends
for some estimators.
Furthermore W97 studied the normalization estimates, and concluded that
the serious discrepancies between the LFs of local and distant galaxies 
is not attributed to the difference of the estimators used in the 
analyses.

Now further questions arise after W97.
The tests of W97 were restricted to the Schechter form LF.
They considered, for example, the bias in the faint-end slope estimation, 
and concluded that even for the spatially homogeneous samples, 
$1/V_{\rm max}$ estimator gives biased results.
It is often claimed that the faint-end overestimation of the 
$1/V_{\rm max}$ estimator is caused by the density inhomogeneity of
the Local Supercluster (e.g. Efstathiou 1996).
Thus if galaxies are homogeneously distributed, 
the estimator is expected to give the correct value.
If any subtler problem dwells in the slope estimation, further extensive
experiments are required.
They also mentioned the binning size selection.
For the analysis of the recent very high-redshift data, data sparseness
should be considered properly.

Recently the LF of galaxies at extremely high redshift has become 
available with the aid of large telescope facilities and improved detectors.
Added to this, redshift surveys have entered upon a new phase by
development of the photometric redshift technique.
The technique requires much lower observational cost than 
the spectroscopic survey,
and is suitable for the analyses of the deep photometric data like the 
Hubble Deep Field (HDF; Williams et al. 1996).
Though some problems are inherent in the technique and in the faint source 
finding itself (Ferguson 1998), vast advances have been produced by the 
method (e.g. Furusawa et al. 2000).
The intermediate--high redshift results are, however, still controversial 
with each other.

To settle down these problems, reliable and robust analyses of the LF
are required.
In this paper, we examined and made practical improvements 
for these estimation methods\footnote{Numerical calculations in this paper 
are based on the public software package for cosmological study 
written by one of the authors (KY).
The C library can be downloaded from 
http://www.kusastro.kyoto-u.ac.jp/$\tilde{~}$kohji/research/libcosm/.}.
Considering the complicated understanding of the evolution of galaxy 
population, we concentrated our discussions on the 
nonparametric methods without any assumed functional forms for the LF.
Besides we restricted our concerns only to the methods which
use the whole sample.
We used the mock catalog generated from various shapes of the probability 
density function (namely the LF).
As we noted above, the density inhomogeneity is a basic property
of the galaxy distribution.
First we tested how accurately these methods reproduce the input density 
function, by using spatially homogeneous mock catalogs with varying 
sample size.
Next, we examined the estimators by using mock catalogs with a dense 
cluster and with a large void.
We also used the  mock 2dF catalog prepared by Cole et al. (1998) in this 
study.
After checking the reliability of each method, we finally 
applied the methods to the photometric redshift catalog prepared 
by Fern\'{a}ndez-Soto et al. (1999) (FLY99) and studied the evolution of the LF at the very large redshift.

This paper is organized as follows:
in Section \ref{sec:lfreview} we review and discuss the methods and our 
extensions.
Section \ref{sec:simulation} is devoted to the tests for the performance 
of these methods by mock catalogs.
We apply the methods to the photometric redshift catalog and discuss the
LF evolution in section \ref{sec:hdf}.
Our summary and conclusions are presented in section \ref{sec:conclusion}.
We briefly introduce the statistical model selection criterion which we used
in our discussions in Appendix \ref{sec:aic}.

\section{NONPARAMETRIC METHODS FOR ESTIMATING LUMINOSITY FUNCTION}
\label{sec:lfreview}

Before we discuss each method, we define some fundamental quantities.
Let $M$ : absolute magnitude, $m$ : apparent magnitude, and 
$\dlz$ : luminosity distance in unit of [Mpc] corresponding to redshift $z$.
Then 
\begin{eqnarray}
  M = m - 5 \log \dlz - 25 - K(z)\, ,
\end{eqnarray}
where $K(z)$ is the $K$-correction.
Here $\log \equiv \log_{10}$.
We use the following notation unless otherwise stated: $\phi (M)$ : 
the luminosity function [${\rm Mpc^{-3}\,mag^{-1}}$], 
$N_{\rm obs}$ : number of detected galaxies in the survey.
When we use stepwise estimators, we must select the optimal binning size to 
suppress the statistical fluctuation (Sturges 1926; Beers 1992 and 
references therein; Sakamoto, Ishiguro, \& Kitagawa 1986; Heyl et al. 1997).
We used Akaike's information criteria (AIC: Akaike 1974) in order to 
select the optimal binning size with the least loss of information 
(Takeuchi 1999; for general discussion, see Sakamoto et~al. 1986).

\subsection{Schmidt--Eales ($1/V_{\rm max}$) Method}\label{sec:vmax}

The method to construct the LF we will discuss here was 
originally proposed by Schmidt (1968)
and well known as the $1/V_{\rm max}$ method.
Eales (1993) developed it further to trace the evolution with redshift.
Cowie et al. (1996) used this estimator in an integral form.

We consider the absolute magnitude and redshift range
\begin{eqnarray}
  \left\{\begin{array}{ll}
      \Ml \leq M \leq \Mu \, \\
      \zl \leq z \leq \zu
    \end{array}
  \right.
\end{eqnarray}
with a survey solid angle $\Omega$ and upper and lower limiting apparent
magnitude, $m_{\rm u}$ and $m_{\rm l}$.
Then we have
\begin{eqnarray}
  \int^{\Mu}_{\Ml} \phi(M) \pd M &=& \sum^{\nobs}_{i=1} 
  \dfrac{1}{\vmax {i}}\, , \\
  \vmax{i} &\equiv& \int_\Omega \int^{\zmax{i}}_{\zmin{i}}
  \dfrac{\pd^2 V}{\pd \Omega \pd z}\, \pd z \pd \Omega \, , 
\end{eqnarray}
where $\zmax{i}$ and $\zmin{i}$ are the upper and lower redshift limits that 
a galaxy with the absolute magnitude $M_i$ can be detected in the survey.
We note that
\begin{eqnarray}
  \zl \leq \zmin{i} < \zmax{i} \leq \zu \, .\nonumber
\end{eqnarray}
Defining $z(M,m)$ to be the redshift that a galaxy with the absolute magnitude 
$M$ is observed as an object with the apparent magnitude $m$,
we get
\begin{eqnarray}
  \zmax{i} &=& \min{\{\zu, z(M_i, m_{\rm u})\}} \, , \\
  \zmin{i} &=& \max{\{\zl, z(M_i, m_{\rm l})\}}\, .
\end{eqnarray}
Actually, both $\zmax{i}$ and $\zmin{i}$ depend on the galaxy spectral energy 
distributions (SEDs).
Thus we must account for the $K$-correction when 
calculating the $1/\vmax{i}$.

Felten (1976) proved that the Schmidt $1/V_{\rm max}$ estimator is unbiased, 
but does not yield a minimum variance.
He also proved that the ``classical estimator $N/V$'', which is different 
from the $1/V_{\rm max}$ estimator, is biased.
Willmer (1997) gave a comment that Felten (1976) had shown this 
estimator to be biased, but it is not exact.
A complication of the terminology may have led to such a comment.

\subsection{Efstathiou--Ellis--Peterson (EEP) Method}\label{sec:eep}

In this subsection, we consider the stepwise maximum likelihood method
introduced by EEP.
Since the estimator of EEP method completely cancels the density information, 
this method requires an independent estimation of the galaxy density.
The EEP method uses the form of the LF
\begin{eqnarray}
  \phi (M) = \sum^{K}_{k=1} \phi_k  W(M_k - M)\, .
\end{eqnarray}
The window function  $W(M_\ell - M)$ is defined by
\begin{eqnarray}
  W(M_\ell - M) \equiv 
  \left\{
    \begin{array}{@{\,}ll}
      1 & {\rm for}\; M_\ell - \dfrac{\Delta M}{2} \leq M \leq M_\ell 
      + \dfrac{\Delta M}{2}\;, \\
      0 & {\rm otherwise} \; .
    \end{array}
  \right.
\end{eqnarray}
According to EEP \footnote{
Koranyi \& Strauss (1997) have pointed out that the discreteness 
of the assumed LF causes a systematic error in the estimation.
In order to avoid this effect, we can use the linear extrapolated form.
We do not discuss it further in this paper (see the appendix of Koranyi \& 
Strauss 1997).}, the likelihood function is
\begin{eqnarray}
  {\cal L} (\{ \phi_k \}_{k=1, \cdots ,K}| \{ M_i \}_{i=1, \cdots ,\nobs})
  = \prod^{\nobs}_{i=1}
  \dfrac{\sum^{K}_{\ell=1}W(M_\ell - M_i)\phi_\ell}{\sum^{K}_{\ell=1}\phi_\ell 
  H(M_{\rm lim}(z_i) - M_\ell) \Delta M}\;,
\end{eqnarray}
\begin{eqnarray}
  H (M_{\rm lim}(z_i) - M) \equiv
  \left\{
    \begin{array}{@{\,}ll}
     1 & M_{\rm lim}(z_i) - \Delta M/2 > M \\
     \dfrac{M_{\rm lim}(z_i) - M}{\Delta M} + \dfrac{1}{2} &
     M_{\rm lim}(z_i) - \Delta M/2 \leq M < 
     M_{\rm lim}(z_i) + \Delta M/2 \\
     0 & M_{\rm lim}(z_i) + \Delta M/2 \leq M
    \end{array} 
  \right. 
\end{eqnarray}
where $M_{\rm lim}(z_i)$ is the absolute magnitude corresponding to the 
survey limit $m_{\rm lim}$ at redshift $z_i$.

The logarithmic likelihood is expressed as
\begin{eqnarray}
  \ln {\cal L} =
  \sum^{\nobs}_{i=1} \left[
  \sum^{K}_{\ell=1}W(M_\ell - M_i) \ln \phi_\ell - 
  \ln \left\{ \sum^{K}_{\ell=1}\phi_\ell H(M_{\rm lim}(z_i) - M_\ell)\Delta M
  \right\} 
  \right] \; .
\end{eqnarray}
Hence, the likelihood equation becomes
\begin{eqnarray}
  \pa{\ln {\cal L}}{\phi_k} =
  \sum^{\nobs}_{i=1} \frac{W(M_k - M_i)}{\phi_k} - 
  \sum^{\nobs}_{i=1} 
  \dfrac{H(M_{\rm lim}(z_i) - M_k) 
  \Delta M}{\sum^{K}_{\ell=1}\phi_\ell H(M_{\rm lim}(z_i) - M_\ell) \Delta M} = 0
\end{eqnarray}
and it reduces to 
\begin{eqnarray}\label{eq:eepphi}
  \phi_k \Delta M = 
  \dfrac{\sum^{\nobs}_{i=1} 
  W(M_k - M_i)}{\sum^{\nobs}_{i=1} 
  \dfrac{H(M_{\rm lim}(z_i) - M_k)}{\sum^{K}_{\ell=1}
  \phi_\ell H(M_{\rm lim}(z_i) - M_\ell) \Delta M}}\; .
\end{eqnarray}
This equation can be solved by iteration, and we obtain the maximum likelihood
estimator $\phat = \{\phat_k\}_{k=1, \cdots, K}$.

As for the normalization of the LF, some estimators have been proposed.
We use the following estimator of the mean galaxy density, $n$, 
which was used by EEP:
\begin{eqnarray}\label{eq:eepnorm}
  n = \dfrac{1}{V} \sum^{\nobs}_{i=1}\dfrac{1}{\Psi (z_i)}\, ,
\end{eqnarray}
where $V$ is the maximum volume defined by the largest redshift in the 
sample, and $\Psi(z)$ is the selection function, defined by
\begin{eqnarray}\label{eq:eepselfunc}
  \Psi (z) \equiv \cfrac{\displaystyle \mathstrut 
    \int^{M_{\rm lim}(z)}_{-\infty} \phi(M)\; \pd M}
  {\displaystyle \mathstrut 
    \int^{\infty}_{-\infty} \phi(M) \;\pd M}\, .
\end{eqnarray}
For further discussions about other normalization estimators, see
the appendix of Davis \& Huchra (1982).
By combining eqs. (\ref{eq:eepphi}) and (\ref{eq:eepnorm}), we get 
the final results.
As Strauss \& Willick (1995) pointed out, at large redshift where
the selection function eq. (\ref{eq:eepselfunc}) is small, 
the estimator eq.~(\ref{eq:eepnorm}) becomes noisy.
Therefore in practice a certain cutoff should be introduced in redshift.

\subsection{Cho{\l}oniewski Method}\label{sec:choloniewski}

Here we discuss the method for estimating the LF developed by C86.
The advantage of the method is that we can obtain the density
and the shape of the LF simultaneously, and can easily examine the galaxy 
density evolution with redshift.
The method explained here is an extended version applicable for 
the sample with a cosmological scale.
In this subsection, we use indices $i,\;j$ as the labels of the cells.

We again consider the absolute magnitude and redshift range
\begin{eqnarray*}
  \left\{\begin{array}{ll}
      \Ml \leq M \leq \Mu \, \\
      \zl \leq z \leq \zu
    \end{array}
  \right.
\end{eqnarray*}
with a survey solid angle $\Omega$ and survey limiting magnitude $m_{\rm lim}$.
And let $n(\vect{r})$ [${\rm Mpc^{-3}}$], the number density of 
galaxies in the neighborhood of the position $\vect{r}$.
If we define $V$ as the total comoving volume under consideration, i.e. 
\begin{eqnarray}
  V = \int_\Omega \int^{\zu}_{\zl} \dfrac{\pd^2 V}{\pd \Omega \pd z}
  \pd z \pd \Omega\, ,
\end{eqnarray}
then it leads to the following expression for the mean number density as
\begin{eqnarray}
  \bar{n} = \dfrac{N}{V}\,,
\end{eqnarray}
where $N$ is the total number of galaxies within the 
redshift range $\zl \leq z \leq \zu$.
Here we adopt a statistical model: on the absolute magnitude--position space
($M$--$\vect{r}$ space) the galaxy distribution is $f(M,\vect{r})$, and 
the probability that we find $k$ galaxies in a volume element $\pd M \pd V$ 
at $(M, \vect{r})$, $P_k$, is described as a Poisson distribution:
\begin{eqnarray}
  P_k &=& \dfrac{e^{- \lambda} \lambda^k }{k !}\, ,\\
  \lambda &=& \dfrac{1}{\bar{n}} \,
  f(M,\vect{r})\,\pd M \pd V \, .
\end{eqnarray}
Here
\begin{eqnarray}
  \phi(M) &=& \dfrac{1}{V} \int_\Omega \int^{\zu}_{\zl} f(M,\vect{r}) 
  \dfrac{\pd^2 V}{\pd \Omega \pd z} \pd z \pd \Omega\, , \\
  n(\vect{r}) &=& \dfrac{1}{V} \int^{\Mu}_{\Ml} f(M,\vect{r}) \pd M\, .
\end{eqnarray}
If we apply an assumption that the random variables $M$ and $\vect{r}$ are 
independent, i.e. $ f(M,\vect{r}) = \psi(M)\nu(\vect{r})$,
then we obtain
\begin{eqnarray}
  \lambda &=& \dfrac{1}{\bar{n}}\,\psi(M) \,\pd M \,\nu (\vect{r}) \pd V \, .
\end{eqnarray}
We integrate $\lambda$ over the spherical shell at redshift $z$ and 
divide the  $M$--$z$ plane into small rectangular cells such that
\begin{eqnarray}
  \begin{array}{ll}
      M_i \leq M \leq M_{i+1} = M_i + \Delta M  & (i = 1, \cdots , I), \\
      z_j \leq z \leq z_{j+1} = z_j + \Delta z & (j = 1, \cdots ,J).
    \end{array}
\end{eqnarray}
Now we see that the problem is to estimate the intensity parameter 
$\lambda_{ij}$ inhomogeneously defined on the  $M$--$z$ plane 
(see Figure~\ref{fig:choloniewski}).
The probability of finding $k_{ij}$ galaxies in the cell $(i,j)$, $P_{k_{ij}}$,
is 
\begin{eqnarray}
  P_{k_{ij}} = \dfrac{e^{-\lambda_{ij}} \lambda_{ij}^{k_{ij}}}{k_{ij}!} \,,
\end{eqnarray}
which is characterized by the parameter 
\begin{eqnarray}
  \lambda_{ij} =
  \int^{M_{i+1}}_{M_i} \int_{V[z_j,z_{j+1}]} 
  \lambda
  \equiv \dfrac{1}{\bar{n}} \, \psi_i \Delta M \, \nu_j V_j \,
\end{eqnarray}
where $V[z_j,z_{j+1}]$ is the comoving volume between redshifts $z_j$ and
$z_{j+1}$.
Here 
\begin{eqnarray}
  \psi_i &\equiv& \frac{1}{\Delta M}\int^{M_{i+1}}_{M_i} \psi (M)\, \pd M \,,\\
  \nu_j &\equiv& \frac{\Omega}{V_j} \int^{z_{j+1}}_{z_j} \nu(\vect{r})
    \dfrac{\pd V}{\pd z} \,\pd z\, , 
\end{eqnarray}
and
\begin{eqnarray}
  V_j &\equiv& \Omega \int^{z_{j+1}}_{z_j} \dfrac{\pd V}{\pd z} 
    \,\pd z\, .
\end{eqnarray}
The likelihood is given by
\begin{eqnarray}
  {\cal L} = {\prod_{(M_i, z_j) \in S}} \dfrac{e^{-\lambda_{ij}} 
    \lambda_{ij}^{k_{ij}}}{k_{ij} !} \,,
\end{eqnarray}
and we obtain the log likelihood
\begin{eqnarray}\label{eq:chlike}
  \ln {\cal L} = {\sum_{(M_i, z_j) \in S}}
  \left\{ k_{ij} \ln \lambda_{ij} - \lambda_{ij} - \ln k_{ij} ! \right\} \,,
\end{eqnarray}
where $S$ stands for the subset of the $M$--$z$ plane surrounded with
$\Mu$, $\Ml$, $\zu$, $\zl$, and the curve ${\cal C}$ defined by the 
selection line
\begin{eqnarray}
  M + 5\log \dlz + K(z) + 25= m_{\rm lim} \,.
\end{eqnarray}
We define the following quantities:
\begin{eqnarray}
  i_{\rm max}(j) \equiv \min{\{I, i_S(j)\}}\, , 
  \; j_{\rm max}(i) \equiv \min{\{J, j_S(i)\}}\, ,
\end{eqnarray}
where 
$M_{i_S(j)} \equiv \{M : {\cal C} \cap \{(M, z) : z = z_j \}\}$,
and $z_{j_S(i)} \equiv \{z : {\cal C} \cap \{(M, z) : M = M_i \}\}$ 
(see Figure~\ref{fig:choloniewski}).
Using these notations, we can reduce eq.~(\ref{eq:chlike}) as
\begin{eqnarray}
  \ln {\cal L} &=& \sum^I_{i = 1} \sum^{j_{\rm max}(i)}_{j = 1} 
  \left\{ k_{ij} \ln \lambda_{ij} - \lambda_{ij} - \ln k_{ij} ! \right\} 
  \,,\nonumber \\
  &=& \sum^J_{j = 1} \sum^{i_{\rm max} (j)}_{i = 1}  
  \left\{ k_{ij} \ln \lambda_{ij} - \lambda_{ij} - \ln k_{ij} ! \right\} \,.
\end{eqnarray}
The maximum likelihood estimates (MLEs) $(\hat{\psi}_0,\cdots, 
\hat{\psi}_i, \hat{\nu}_0, \cdots , \hat{\nu}_J, \hat{\bar{n}})$ 
are the set of solutions which maximizes ${\cal L}$.
They can be obtained, in practice, by setting the following equations to zero:
\begin{eqnarray}
  \pa{\ln {\cal L}}{\psi_i} &=& \sum^I_{s=1}
  \sum^{t_{\rm max}(s)}_{t=1} \left( k_{st} \frac{1}{\lambda_{st}}
  \pa{\lambda_{st}}{\psi_i} - \pa{\lambda_{st}}{\psi_i} \right) \nonumber \\
  &=& \sum^{t_{\rm max}(i)}_{t=1} 
  \left( \frac{k_{it}}{\psi_i} - \frac{1}{\bar{n}} \Delta M \, \nu_t V_t 
  \right) = 0 \, , \\
  \pa{\ln {\cal L}}{\nu_j} &=& \sum^J_{t=1}
  \sum^{s_{\rm max}(t)}_{s=1} \left( k_{st} \frac{1}{\lambda_{st}}
  \pa{\lambda_{st}}{\nu_j} - \pa{\lambda_{st}}{\nu_j} \right) \nonumber \\
  &=& \sum^{s_{\rm max}(j)}_{s=1} 
  \left( \frac{k_{sj}}{\nu_j} - \frac{1}{\bar{n}} \psi_s \, \Delta M V_j 
  \right) = 0\, .
\end{eqnarray}
Thus we have a set of equations which are referred to as likelihood 
equations.
\begin{eqnarray}
  \frac{\Delta M}{\bar{n}} \psi_i \sum^{t_{\rm max}(i)}_{t=1} 
  \nu_t V_t  &=& \sum^{t_{\rm max}(i)}_{t=1} k_{it} \, , 
  \label{eq:chphimle}\\
  \frac{\Delta M}{\bar{n}} \nu_j V_j \sum^{s_{\rm max}(j)}_{s=1} 
  \psi_s &=& \sum^{s_{\rm max}(j)}_{s=1} k_{sj} \, .
  \label{eq:chnmle}
\end{eqnarray}
These equations are solved by iterative procedure.
At this stage, these solutions obtained here are not exactly the MLEs 
themselves, but relative values. 
We need one more step to obtain absolute values. 
We denote the relative solutions by `$\tilde{~}$' and exact MLEs 
by `$\hat{~}$'.
From eqs. (\ref{eq:chphimle}) and (\ref{eq:chnmle}), we have
\begin{eqnarray}
  \tilde{\psi}_i &=& 
  \frac{\bar{n}}{\Delta M}\, 
  \dfrac{\sum^{t_{\rm max}(i)}_{t=1} k_{it}}{\sum^{t_{\rm max}(i)}_{t=1} 
    \tilde{\nu}_t V_t} = \dfrac{\sum^{t_{\rm max}(i)}_{t=1} k_{it}}{
    \sum^{t_{\rm max}(i)}_{t=1}
    \dfrac{\sum^{s_{\rm max}(t)}_{s=1} k_{st}}{\sum^{s_{\rm max}(t)}_{s=1} 
      \tilde{\psi}_s}} \, ,\\
  \tilde{\nu}_j V_j &=& 
  \frac{\bar{n}}{\Delta M}\, 
  \dfrac{\sum^{s_{\rm max}(j)}_{s=1} k_{sj}}{\sum^{s_{\rm max}(j)}_{s=1} 
    \tilde{\psi}_s}  = \dfrac{\sum^{s_{\rm max}(j)}_{s=1} k_{sj}}{
    \sum^{s_{\rm max}(j)}_{s=1}
    \dfrac{\sum^{t_{\rm max}(s)}_{t=1} k_{st}}{\sum^{t_{\rm max}(s)}_{t=1} 
      \tilde{\nu}_t V_t}}\, .
\end{eqnarray}
Then we properly normalize these solutions. 
Clearly it follows that
\begin{eqnarray}\label{eq:chnormalization}
  \sum_{(M_i, z_j) \in S} \lambda_{ij} = 
  \Delta M \sum_{(M_i,z_j) \in S} \hat{\psi}_i \;\hat{\nu}_j V_j = 
  N_{\rm obs}\, .
\end{eqnarray}
If we set $\tilde{\psi}_i \,\tilde{\nu}_j = w \,\hat{\psi}_i \,\hat{\nu}_j$, 
then we straightforwardly obtain the numerical factor $w$ by 
eq.~(\ref{eq:chnormalization}):
\begin{eqnarray}
  w = \dfrac{N_{\rm obs}}{\Delta M \sum_{(M_i,z_j) \in S} (\tilde{\psi}_i 
    \;\tilde{\nu}_j V_j)}\, .
\end{eqnarray}
Now we obtain the LF $\phi(M)$ and density $n(z)$ as
\begin{eqnarray}
  \phi(M_i) &=& \dfrac{1}{V} \hat{\psi}_i 
  \sum^J_{j=1} \hat{\nu}_j V_j\, , \\
  n(z_i) &=& \dfrac{\Delta M}{V} \hat{\nu}_j V_j 
  \sum^I_{i=1} \hat{\psi}_i \, .
\end{eqnarray}


\subsection{Lynden-Bell--Cho{\l}oniewski--Caditz--Petrosian (LCCP) 
  Method}\label{sec:lccp}

In this section, we discuss the method originally introduced by
Lynden-Bell (1971) as the `$C^-$ method'.
The estimator of this method is an analog of the Kaplan--Meier estimator 
used for censored data analyses, like survival analysis (e.g. Feigelson 
\& Nelson 1985; Feigelson 1992; Babu \& Feigelson 1996; for reference of 
survival analysis itself, see e.g. Kleinbaum 1996).
The method may be the most natural application of the nonparametric 
statistics to the problem (e.g. Petrosian 1992).
The rederived version of the method by C87 was improved
so that it could estimate the LF and density evolution of galaxies at 
the same time.
In addition, the derivation of the estimator was much simplified.
The original method was invented to estimate the cumulative LF as 
a step function, 
thus the differential LF was described as a weighted sum of Dirac's 
$\delta$-function.
But obviously this form is not practical, and C87 suggested 
to smooth the LF.
In modern statistics, the kernel estimator is used in the problem of 
nonparametric density estimation (Silverman 1986; Lehmann 1999).
The kernel is a smooth function which is used as a substitute of 
the delta function, in order to keep the estimated density function smooth.
This improvement was introduced to the LF estimation problem by CP93, 
and used for a photometric redshift catalog 
by Subba Rao et al. (1996)\footnote{But we note that Subba Rao et al.'s eq.(6) 
erroneously includes an extra exponential.}.

We unify these improvements, and show the practically convenient
calculation here, which we call the `LCCP method' after the names 
of the above contributors.
We use the same notations for luminosity function, galaxy number density,
distribution of galaxies, etc., and we consider the same absolute magnitude
and redshift ranges as in section \ref{sec:choloniewski}.
But we must note that, in this subsection, indices represent the labels
of galaxies.
This method is completely free of binning procedure.

For the later discussion, we suppose that the galaxies are ordered as
$M_k \leq M_{k+1}$.
In the LCCP method, the independence assumption is also adopted for 
$M$ and $z$, which leads to the expression
\begin{eqnarray}
  f(M,z) = \psi(M)\nu(z)\, .\nonumber
\end{eqnarray}
The empirical distribution (distribution of observational data) is 
expressed as
\begin{eqnarray}
  \fobs = \sum^{\nobs}_{k=1} \delta (M-M_k, z-z_k)\, ,
\end{eqnarray}
again $\nobs$ is the observed sample size, and let
\begin{eqnarray}
  \psi (M) &=& \sum^{\nobs}_{i=1} \psi_i \;\delta (M-M_i)\, , \\
  \nu (z) &=& \sum^{\nobs}_{j=1} \nu_j \;\delta (z-z_j)\, .
\end{eqnarray}
Then the empirical distribution is
\begin{eqnarray}
  \fobs &=& \sum^{\nobs}_{i=1} \psi_i \;\delta (M-M_i)\, 
  \sum^{\nobs}_{j=1} \nu_j \;\delta (z-z_j)\; \chis \nonumber\\
  &=& \sum_{(i,j) \in S} \psi_i\nu_j \delta(M-M_i)\delta(z-z_j)\, .
  \label{eq:empdist}
\end{eqnarray}
Here, $\chis$ is the characteristic function of the set $S$
defined as
\begin{eqnarray}
  \chis \equiv 
  \left\{
    \begin{array}{ll}
      0 & (M, z) \notin S\, , \\
      1 & (M, z) \in S\,.
    \end{array}
  \right. \nonumber
\end{eqnarray}
In the following discussions, the quantities $M_{\rm max}(j)$ and 
$z_{\rm max}(i)$ are defined as
\begin{eqnarray}
  M_{\rm max}(j) \equiv \min{\{\Mu, M_{S(j)}\}}\, , 
  \; z_{\rm max}(i) \equiv \min{\{\zu, z_{S(i)}\}}\, ,
\end{eqnarray}
where 
$M_{S(j)} \equiv \{M : {\cal C} \cap \{(M, z) : z = z_j \}\}$,
and $z_{S(i)} \equiv \{z : {\cal C} \cap \{(M, z) : M = M_i \}\}$.
Though they look like those used in the subsection 
\ref{sec:choloniewski}, we note again that the indices are of galaxies.
These are schematically described in Figure~\ref{fig:lccp}.
Integration of eq. (\ref{eq:empdist}) over the interval $[\Ml, \Mu],\;
[z_k - \varepsilon , z_k + \varepsilon]$ ($\varepsilon > 0$) gives
\begin{eqnarray}
  &&\int^{\Mu}_{\Ml} \int^{z_k + \varepsilon}_{z_k - \varepsilon} 
  \sum^{\nobs}_{\ell=1}  \delta (M-M_\ell, z-z_\ell) \pd z \pd M = 1 
\nonumber \\
  &=& \int^{\Mu}_{\Ml} \int^{z_k + \varepsilon}_{z_k - \varepsilon} 
  \sum_{(i,j) \in S} \psi_i\nu_j \delta(M-M_i)\delta(z-z_j) \pd z \pd M 
  \nonumber \\
  &=& \nu_k \sum^{M_i < M_{{\rm max} (k)}}_{i=1} \psi_i\, \nonumber
\end{eqnarray}
therefore
\begin{eqnarray}
   \nu_j \sum^{M_i < M_{{\rm max} (j)}}_{i=1} \psi_i = 1\; , \quad \quad 
   (j = 1,\cdots , \nobs) \, . \label{eq:lccpcoefi}
\end{eqnarray}
Similarly, integration over the $[M_k - \varepsilon , M_k + \varepsilon],\; 
[\zl, \zu]$ gives
\begin{eqnarray}
   \psi_i \sum^{z_j < z_{{\rm max} (i)}}_{j=1} \nu_j = 1\; , \quad \quad 
   (i = 1,\cdots , \nobs)\, . \label{eq:lccpcoefj}
\end{eqnarray}
Formally, we can obtain $\{\psi_i \}_{i = 1,\cdots , \nobs}$ and
$\{\nu_j \}_{j = 1,\cdots , \nobs}$ by solving the eqs. (\ref{eq:lccpcoefi}) 
and (\ref{eq:lccpcoefj}), and the estimates of the real galaxy 
distribution, $f(M, z)$, as
\begin{eqnarray}
   f(M,z) = \sum^{\nobs}_{i=1} \sum^{\nobs}_{j=1} 
   \psi_i\;\delta(M-M_i)\; \nu_j\;\delta(z-z_j) \, .
\end{eqnarray}
Thus 
\begin{eqnarray}
  \phi (M) &=& \frac{1}{V} \int^{\zu}_{\zl} f(M,z) \pd z = 
  \dfrac{1}{V} \sum^{\nobs}_{i=1} \psi_i\;\delta (M-M_i) 
  \sum^{\nobs}_{j=1} \nu_j \, , \\
  n(z) &=& \dfrac{\pd z}{\pd V} \int^{\Mu}_{\Ml} f(M,z) \pd M =
  \dfrac{\pd z}{\pd V} \sum^{\nobs}_{j=1} \nu_j \delta (z-z_j)
  \sum^{\nobs}_{i=1} \psi_i \;,
\end{eqnarray}
where $V$ is the volume considered.
The total number of galaxies $N$ is 
\begin{eqnarray}
  N = \sum^{\nobs}_{i=1} \psi_i\; \sum^{\nobs}_{j=1} \nu_j \,.
\end{eqnarray}
These solutions are MLEs as discussed in C87.

In spite of the clarity of the derivation, it is, actually, not an easy
task to solve the eqs. (\ref{eq:lccpcoefi}) and (\ref{eq:lccpcoefj}) 
numerically if the data size $\nobs$ is large.
Thus we use the usual $C$ estimator together, in order to calculate 
the Cho{\l}oniewski's coefficients more easily.
The Lynden-Bell's $C^-$-function, $C^- (M_k)$, is the number of galaxies 
in the region
\begin{eqnarray}
  \left\{
    \begin{array}{l}
      \Ml \leq M < M_k \, , \\
      \zl \leq z \leq z_{{\rm max} (k)} \, .
    \end{array}
  \right.
\end{eqnarray}
Let
\begin{eqnarray}
  C_k \equiv C^- (M_k)\, , \quad k = 1, \cdots , \nobs\, .
\end{eqnarray}
Then, using eqs. (\ref{eq:lccpcoefi}) and (\ref{eq:lccpcoefj}),
we have
\begin{eqnarray}
  C_{k} + 1 = \sum^k_{i=1} \psi_i \sum^{z_j < z_{{\rm max} (k)}}_{j=1} \nu_j = 
  \sum^{k}_{i=1} \dfrac{\psi_{i}}{\psi_{k}} \, .
\end{eqnarray}
and we obtain the following recursion relation:
\begin{eqnarray}
  \psi_{k+1} = \dfrac{C_k + 1}{C_{k+1}} \psi_k \, .\label{eq:lccprecursion}
\end{eqnarray}
Thus, the distribution function of $M$ (cumulative LF), $\Phi(M)$, is 
\begin{eqnarray}
  \Phi (M) \propto \sum^{M_k < M}_{k=1} \psi_k 
  = \psi_1 \prod^{M_k < M}_{k=1} \dfrac{C_k +1}{C_k}\, .
\end{eqnarray}
In the real procedure, we set $(C_1 +1)/C_1 = 1$, so the product in the
above equation begins with $k=2$.
We can prove the second step of the above equation by mathematical induction.
This is equivalent to the Lynden-Bell's solution (C87).
We obtain the weight $\{\psi_i \}_{i=1, \cdots, \nobs}$ by eq. 
(\ref{eq:lccprecursion}), and we are able to calculate the density weight
 $\{\nu_j \}_{j=1, \cdots, \nobs}$ by eq. (\ref{eq:lccpcoefj}).

As we mentioned above, the weighted sum of the $\delta$-function is 
not a practically useful form, and random fluctuation would be serious 
in the region where the data points are sparse.
Therefore, the kernel estimator, which is often used in modern nonparametric 
density estimation, was introduced by CP93.
This estimator is simply obtained by replacing the $\delta$-function with
a smooth kernel function $\kappa$ as
\begin{eqnarray}
  f(M,z) = \sum^{\nobs}_{i=1} \sum^{\nobs}_{j=1} \psi_i\; \nu_j \,
  \dfrac{1}{h_M h_z} \kappa \left( \dfrac{M-M_i}{h_M}\right) 
  \kappa \left(\dfrac{z-z_j}{h_z} \right)\, .
\end{eqnarray}
The minimum value of the `smoothing scale' $h$ is restricted by the 
observational uncertainty, which was used by Subba Rao et al. (1996), 
but it does not provide sufficient smoothing in general (CP93).
The optimal value of $h_M$ or $h_z$ may be estimated as
\begin{eqnarray}
  h_M \sim \max{ \{ M_{i+1} - M_{i}\}}_{i = 1, \cdots , \nobs}\, , \; 
  h_z \sim \max{ \{ z_{j+1} - z_{j}\}}_{j = 1, \cdots , \nobs}\, .
\end{eqnarray}
It is obvious that the larger the data size $\nobs$ is, the smaller
the smoothing scale becomes.
Furthermore, CP93 discussed the effect of the kernel shape on the estimates.
Now it is known that the best shape of the kernel is parabolic, so-called
the Epanechnikov kernel (Epanechnikov 1969), because it
gives the minimum variance (Lehmann 1999; van Es 1991):
\begin{eqnarray}
  \kappa ( x ) = \frac{3}{4}(1-x^2) .
\end{eqnarray}
It should be noted that, in principle, the kernel estimator is asymptotically 
biased, i.e. the expectation value is slightly different from the 
true value even if the sample size is large.

\section{TEST OF THE METHODS BY SIMULATION}\label{sec:simulation}

\subsection{Numerical Examination with Mock Catalogs}\label{subsec:mock}

The validity of the estimation methods of the LF is often
examined by mathematical statistics.
For example, their statistical unbiasedness and statistical convergence
were discussed in many early works (e.g. Felten 1976).
However, quantitative evaluation frequently appears to be difficult by 
such approach, and numerical examination is quite important.
Jackson (1974) used numerical experiments, as well as the analytical 
error estimation by Fisher's information matrix (see Stuart, Ord, \& 
Arnold 1999), in the study of quasar LF, and EEP also checked the errors of 
their method by Monte Carlo simulations as well as traditional information 
matrix approach.
Mobasher, Sharples, \& Ellis (1993) performed Monte Carlo error estimation
to test the special method developed to construct the LF at a certain 
waveband from the data selected at another wavelength.
Heyl et al. (1997) examined the effect of galaxy clustering to the LF 
estimation by their extended EEP method.
But computer-aided extensive intercomparison between the estimators 
had not been performed until the work of W97.
They discussed the performance of several estimators when the LF is represented
by the Schechter form
\begin{eqnarray}
  \phi (M)\, \pd M = 0.4 \ln 10 \, \phi_{*} 10^{-0.4 (\alpha +1) (M - M_{*})}
  \exp \left( -10^{-0.4(M - M_{*})} \right) \pd M \,, 
\end{eqnarray}
and tested the results in some cases with different
Schechter parameters.
Their main conclusions are as follows:
\begin{enumerate}
\item The STY and $C^-$ methods are the best.
\item The $1/V_{\rm max}$ method gives biased results and tends to give
  higher values for the faint-end slope {\sl even for spatially homogeneous
  samples}.
\item The STY fit tends to underestimate the faint-end slope.
\item The mean densities (normalization of the LF) recovered by most estimators
  are lower than the input values by factors (up to 20 \%). 
\end{enumerate}
Among these, the second one looks most strange, because as we mentioned in 
section \ref{sec:vmax}, Felten has proved mathematically that the 
$1/V_{\rm max}$ estimator is unbiased when the homogeneous assumption holds.
The $1/V_{\rm max}$ method is quite frequently used in the estimation for
the LFs of quasars, clusters of galaxies, etc., and if W97's claim is 
true, some widely accepted conclusions must be significantly changed.
Thus it is necessary to examine the estimators further, not only for the 
Schechter form but for various shapes of the LF, in order to clarify the
trends of the results.

In this section, we test the four estimators discussed in the previous 
section by using simulated mock galaxy samples with a variety of the LFs 
which have the following functional forms:
\begin{enumerate}
\item[A.] Uniform distribution,
\item[B.] Power-law form which increases toward fainter magnitude,
\item[C.] Power-law form which decreases toward fainter magnitude,
\item[D.] Gaussian distribution (with standard deviation 1.67 mag),
\item[E.] Schechter form (steep faint-end slope: $\alpha = -1.6$),
\item[F.] Schechter form (flat faint-end slope: $\alpha = -1.1$),
\end{enumerate}
with magnitude range $M = [-24, -14]\,$ (Figure~\ref{fig:lfforms}).
The first three forms are designed to examine the effect of LF slope for 
estimation, and the form D, Gaussian, is to check the effect of curvature 
of the function.
Power-law LF of the form B appears ubiquitously in various types of objects.
The form C looks apparently unrealistic, but we added this for making 
thorough investigation.
The form D is interesting because approximate Gaussian form is often found 
in the LFs of individual galaxy types.
We applied Box--Muller method (Box \& Muller 1958) to generate Gaussian 
distribution from uniform random number, and von Neumann's 
acception--rejection method to obtain other distributions 
(see Knuth 1998 for details).
We set the sample sizes $\sim 100$ and $\sim 1000$, to study the behavior
of the statistical estimators with galaxy number.
Here `sample size' means the detected number of galaxies after magnitude
selection (observation) procedures.
Therefore the underlying population density for each LF form is different 
from each other.
The estimation of galaxy spatial density is an important part of the
derivation of the LF.
What to be estimated is the total galaxy number including the galaxies
too faint to be observed.
In our simulations, we stochastically produced galaxies according to 
the assumed LF, distribute them in space, calculate their observed flux,
and judge that they could be observed or not.
Therefore, the total number corresponds to the number of Monte Carlo trials.
We fixed the number of trials through one sequence of simulations with 
a certain LF shape and spatial density.

\subsubsection{Mock Catalog with Spatially Homogeneous Distribution}

First we construct a set of mock galaxy samples with spatially homogeneous
distribution in order to investigate the bias trend of the estimators, 
especially for the Schmidt and Eales' $1/V_{\rm max}$.
We set the redshift range up to 0.1, and we adopted the Hubble parameter 
$H_0 = 75\;{\rm km\,s^{-1} Mpc^{-1}}$, $\Omega_0 = 0.2 \;(q_0 = 0.1)$, 
$\lambda_0 = 0$, 
and limiting magnitude $m_{\rm lim} = 13$ mag in the series of simulations.
No $K$-correction is considered here.
We constructed 100 representations for each LF form and sample size, and 
applied the four estimators to each sample.

\subsubsection{Mock Catalog with a Dense Cluster and with a Void}

We, next, investigate the response against density inhomogeneity 
of galaxies.
We consider some extreme cases for clear understanding.
For the case with density enhancement, we constructed a series of mock 
catalogs with a dense spherical clump, to which half of the galaxies belong.
The clump lies at a distance of 0.8 Mpc, and its radius is 0.8 Mpc.
We call the mock catalog the ``cluster sample''.
An example of the spatial configuration of galaxies of a cluster sample is 
described in Figure~\ref{fig:clustersample}.
Then we also constructed a set of the mock catalogs with a large spherical 
void without galaxies.
The void lies at a distance of 0.8 Mpc and its radius is 1.6 Mpc.
We call this mock catalog the ``void sample''.
The overall underlying density of cluster and void samples 
defined in a considered volume is the same as the homogeneous samples for 
each LF shape, i.e. we set the number of Monte Carlo trials the same 
as that of the homogeneous sample for each LF shape.
Therefore, the observed sample size of the cluster sample is larger than 
that of the homogeneous sample, because we put the dense clump in the 
considered volume.
In the case of the void sample, the observed galaxy number is smaller than
that of the homogeneous one.

\subsubsection{Results}

The results for the 1000-galaxy samples are shown in Figure
5 -- 10.
The solid lines represent the input distributions, and the symbols are
the averages of the estimates.
The error bars depict the standard deviations of the mean of the estimates
for 100 representations.
Figures 5a, 5b, and 5c are the results from the spatially homogeneous sample, 
from the cluster sample, and from the void sample, respectively.
This is also the same for Figures 6 -- 10.

At a glance, we see that all estimators give consistent results with each 
other, and we do not find any bias trends in our numerical experiments 
for any LF forms in the case of homogeneous samples.
For cluster samples, the $1/V_{\rm max}$ method yields strongly distorted
estimations, as widely recognized.
The overestimation corresponding to the dense clump clearly appeared
in the $1/V_{\rm max}$ results.
In contrast, the other three estimators were not affected by the dense 
cluster at all.
The estimates appeared to be consistent with each other, 
and showed perfect agreement with the input LFs.
The $1/V_{\rm max}$ method was also affected by the large void, and gave
underestimated results.

Large fluctuations appear at the faint end of the LF, because the number of 
available data points is small, especially in the case of the LF form C and D.
We can obtain statistically stable estimates if the slope is properly steep,
and the more shallow the slope is, the larger the fluctuation becomes.
This is clearly shown in Figure 5 -- 10.

In principle, the error bar of the Cho{\l}oniewski method is larger than those
of the other methods, because the method subdivides the $M - z$ plane both 
in $M$ and $z$.
This procedure enables us to estimate the shape, the normalization, and the 
evolution of the LF at the same time.
On the other hand, this becomes a drawback when the data size is small, 
because the shot noise dominates.
Therefore we cannot expect a firm estimation with the Cho{\l}oniewski method
when the sample size is smaller than 100.

Here we mention the calculation time that each method needed 
for the same sample size.
Because of its algorithmic simplicity, Cho{\l}oniewski method is the fastest
among the four methods.
When we analyze the 1000-mock data, the relative calculation times 
of the $1/V_{\rm max}$, EEP, and LCCP methods normalized with that of 
Cho{\l}oniewski method are 2.76, 2.73, and 1.87, respectively.
This advantage is quite significant when we treat a large sample of 
$\sim 10^{4 - 5}$ galaxies.
We estimate the LFs from large datasets of sample size 250,000 
in Section~\ref{subsec:2df} by the $1/V_{\rm max}$, EEP, and 
Cho{\l}oniewski methods.
The relative calculation times of the $1/V_{\rm max}$ and EEP methods 
normalized with that of Cho{\l}oniewski method are, in this case, 
15.01 and 131.74, respectively.
The EEP method takes longer calculation time because 
it needs more iterations in the procedure than others do.
The $1/V_{\rm max}$ method derives the maximum volume $V_{\rm max}$ for 
each galaxy, and also needs some calculation time.
The LCCP method requires a large stack for data sorting procedure, which 
is a requirement of this method.
Thus we stress that the Cho{\l}oniewski method is most economic from 
the standpoint of practical computing.

Figures 
11 -- 16 are the same as Figures 5 -- 10, except that the data size is 100.
We see it is often not possible to determine the faint end of the LF 
accurately for such small datasets.
The fluctuation became larger than the result of the 1000-sample, but we  
did not find the systematic bias trend from our results.
Thus we conclude that when the galaxy distribution is homogeneous, 
all four estimators provide the consistent and correct results, even 
the $1/V_{\rm max}$ estimator.

\subsection{Mock 2dF Redshift Catalog}\label{subsec:2df}

The Anglo--Australian 2-degree field (2dF) galaxy redshift survey is now
underway\footnote{See http://msowww.anu.edu.au/\~{}colless/2dF for recent 
status.}.
This survey will measure 250,000 redshifts, up to $z \sim 0.2$, and be 
complete to an extinction corrected apparent magnitude of $b_{\rm J} < 
19.45$ mag.
In order to develop statistical methods and faster algorithms for 
the analyses of such large upcoming redshift surveys, Cole et al. (1998) 
prepared an extensive set of mock 2dF catalogs constructed from a 
series of large cosmological $N$-body simulations.
The simulations span a wide range of cosmological models, with various 
values of the density parameters, $\Omega_0$, the cosmological constant, 
$\lambda_0$, and the shape parameter $\Gamma$ and amplitude of the density
fluctuation $\sigma_8$.
The LF is assumed to be a Schechter form with the parameters
reported by APM-Stromlo bright galaxy survey (Loveday et al. 1992), 
$M_{b_{\rm J} *} - 5\log h = -19.5$ mag, $\alpha = -0.97$, 
and $\phi_{*} = 1.4 \times 10^{-2}\,h^3\;{\rm Mpc^{-3}}$.
The $K$-correction is assumed to be canceled by evolutionary correction.

We applied the three methods to the mock 2dF catalog in 
order to see how accurately they can reproduce the true LF when they
are used in the analysis of realistic large redshift surveys.
We did not use the LCCP method for this sample.
When we treat such a large catalog, the advantage of the Cho{\l}oniewski
methods is extremely significant.
We also focused on the difference between the real-space data and 
the redshift-space data which is affected by the redshift distortion.
The redshift distortion causes a scatter in the estimated luminosities of
galaxies.
In this study, we used three mock catalogs, named E1 (Einstein--de Sitter: 
$\Omega_0 = 1, \lambda_0 = 0, \Gamma = 0.5, \sigma_8 = 0.55$), 
L3S ($\Omega_0 = 0.3, \lambda_0 = 0.7, \Gamma = 0.25, \sigma_8 = 1.13$), 
and O3S ($\Omega_0 = 0.3, \lambda_0 = 0, \Gamma = 0.25, \sigma_8 = 1.13$).
The catalogs we selected are all cluster-normalized, i.e. the amplitude
of the initial power spectrum is set to reproduce present abundance
of rich galaxy clusters in the local Universe (e.g. Viana \& Liddle 1996;
Kitayama \& Suto 1997) and $h = \Gamma/\Omega_0$.

We compare the input LF and the estimated LF in Figures~\ref{fig:2dFLF_EdS}, 
\ref{fig:2dFLF_L3S}, and \ref{fig:2dFLF_O3S}.
Figure~\ref{fig:2dFLF_EdS} shows the LF derived from the Einstein--de Sitter
(EdS) Universe, Figure~\ref{fig:2dFLF_L3S} is the LF derived from L3S data, 
and Figure~\ref{fig:2dFLF_O3S} is the LF derived from O3S data.
The left panels in these Figures show the LFs derived from the 
redshift-space data, and the right panels, those from the real-space data.
First, we see that all the estimators provided perfectly
consistent results, and they show an excellent agreement with the input LF.
There are no significant difference between the real- and redshift-space 
datasets.
The slight deviations of $1/V_{\rm max}$ estimates are caused by 
the clustering in the 2dF mock catalog.
Thus we do not have to consider the redshift distortion effect seriously 
when we derive the galaxy LF from such large-volume redshift surveys.
When we use such a large survey, we should rather mention the photometric 
calibration as a more important error source.

\section{APPLICATION TO THE HUBBLE DEEP FIELD}\label{sec:hdf}

Recently some authors claim that the faint-end slope of the LF becomes
steeper with redshift at $z < 1$ (e.g. Ellis et al. 1996; Heyl et al. 1997; 
but see Lin et~al. 1999).
The LFs for some special classes of galaxies such as Lyman-break objects 
(Steidel et al. 1998) or Ly-$\alpha$ emitters (Pascarelle et al. 1999)
are now also available.
We, however, do not have a coherent understanding of the evolution of the LF
and the evolution of the luminosity density, $\rho_{\rm L}$.
At low redshift, Zucca et al. (1997) reported a high normalization LF with 
$\phi_* = 0.020\, h^3 \; {\rm Mpc^{-3}}$, and Ellis et al. (1996) 
obtained $\phi_* = 0.026\, h^3 \; {\rm Mpc^{-3}}$, while Loveday et al. 
(1992) derived $\phi_* = 0.014\, h^3 \; {\rm Mpc^{-3}}$, and Las Campanas 
Redshift Survey result (Lin et al. 1996) is similar to the value of 
Loveday et al.~(1992).
The local value of the LF parameters plays a crucial role in the study of 
galaxy evolution, since it controls the redshift dependence of 
$\rho_{\rm L}$.
Cowie et al. (1999) showed a rather mild evolution of the UV luminosity 
density at $z < 1$ from their surveys.
On the other hand, high redshift LF estimations are also controversial with
each other.
Gwyn \& Hartwick (1996) claimed dramatic changes in the LF from $z=0$ to 
$z \sim 5$, becoming flat between $-24 \leq M_B \leq -15$ for $3 < z < 5$.
On the contrary, SLY97 reported more familiar Schechter form 
with $\alpha = -1.3$ for the LF at $3< z < 4$.
Mobasher et al. (1996) suggested a stronger evolution of the LF.
From a deep multiband photometric survey, Bershady et al.~(1997) gave
a constraint which ruled out Gwyn \& Hartwick's result.

Thus in this section, we apply the four estimators to the 
photometric redshift catalog of the HDF to study the evolution of the
LF shape.
For the observational data, the error estimation is complicated, because
the estimation procedure of the LF involves the magnitude selection, 
weighting, etc.
In such cases, bootstrap resampling analysis is known to be often superior
to classical analytic methods in order to estimate statistical properties
(e.g. Efron \& Tibshirani 1993; Babu \& Feigelson 1996; Davison \& Hinkley 
1997).
Thus we used the bootstrap method for the estimation of the statistical 
uncertainties.
When we perform the bootstrapping, how to generate good random numbers is 
important.
We generated the uniform random number by Mersenne Twister method\footnote{
For recent development, see http://www.math.keio.ac.jp/matumoto/mt.html.}
developed by Matsumoto \& Nishimura (1998).

\subsection{Sample}

We used the photometry and photometric redshift catalog of the HDF
prepared by FLY99.
Their catalog contains 1067 galaxies, with ${\rm AB}(8140) < 26.0$.
The photometric redshifts are derived based on both {\it UBVI}
(F300W, F450W, F606W, and F814W, respectively; Williams et al. 1996) 
obtained by WFPC2, and {\it JHK} obtained by the IRIM camera on the 
Kitt Peak National Observatory 4-m telescope.
The object detection and photometry are performed using SExtractor (Bertin
\& Arnouts 1996).
Details of the procedures are found in FLY99.
In the peripheral region of the WFPC2 image (referred to as zone 2), the 
detection limit is ${\rm AB}(8140) = 26$ mag, and in the inner region 
(zone 1), ${\rm AB}(8140) = 28$ mag.
We restricted our analysis to the inner zone 1 sample.
The solid angle of zone 1 is $3.92 \; {\rm arcmin^2} = 3.32 \times
10^{-7}\; {\rm sr}$.
The sample size is then 946 galaxies.

FLY99 used four spectral templates given by 
Coleman, Wu, \& Weedman (1980) to determine the photometric redshifts.
For ultraviolet wavelengths, the templates are extrapolated by using the
results of Kinney et al. (1993), and for infrared, by the models of 
Bruzual \& Charlot (1993).
Evolutionary corrections are not included in the model spectra to avoid
additional parameter dependence.
According to Coleman et al. (1980), they classified the galaxy spectra into
four categories: 1. Elliptical, 2. Sbc, 3. Scd, and 4. Irr.
We used these labels to set the $K$-corrections.

In principle, the SED must be the same as the templates used in 
FLY99, but for simplicity and comparison with other studies, we used 
the galaxy SED sample compiled by Kinney et al. (1996).
The data of Kinney et al. (1996) have almost the same properties as
the SED templates of FLY99, thus we can use them comfortably.
To construct the $K$-correction function, we first selected the 
sample galaxy SEDs corresponding to the labels of FLY99, and fitted
polynomial functions from 1st order to 6th order.
The order of polynomial fitting was decided by referring to AIC, and
we chose the 5th order.

\subsection{Results and Discussions}

We show the redshift-dependent LF at $I$-band and $B$-band in 
Figures~\ref{fig:HDFLF_I} and \ref{fig:HDFLF_B}, respectively.
The symbols represent the estimated LFs by the four methods.
We show the LFs of the HDF at $0 < z < 0.5$ (106 galaxies), $0.5 < z < 1.0$
(193), $1.0 < z < 1.5$ (204), $1.5 < z < 2.0$ (193), $2.0 < z < 3.0$ (117), 
and $3.0 < z < 6.0$ (109).
The sample is $I$-band selected, and we derived the $B$-band LF by following
the discussion of Lilly et al.~(1995).
We stress that the four different LF estimators give consistent
results for the HDF sample, same as the results for the mock catalogs.

We clearly see the evolutionary trend of the LF with redshift.
But we note that, though we can fit Schechter function, it is not so easy to 
derive the parameters $\alpha$, $M_*$, or $\phi_*$ precisely, because the 
Schechter function is rather smooth and the errors of these characteristic 
parameters are strongly correlated.
These parameters can be easily affected by statistical fluctuations.
We will discuss more details of the $I$- and $B$-band results at each redshift 
range in the following.

\subsubsection{$I$-band LF at $0 < z < 0.5$}

In Figure~\ref{fig:HDFLF_I}, the dotted line represents the local $I$-band LF 
obtained by Metcalfe et al. (1998).
Metcalfe et al. (1998) pointed out a possible upturn of the faint end of 
their multiband LFs, though they took a prudent attitude in concluding 
firmly.
The upturn magnitude $M_I \sim -15 + \log h$ mag (in 
Figure~\ref{fig:HDFLF_I}, $h=0.75$) is in good agreement with that of our 
lowest redshift LF except the normalization.

\subsubsection{Evolution at $B$-band: $0 < z < 0.5$}

We compare the normalization of the LF with other previous results.
Our $B$-band LF shows roughly good agreement with other local LFs.
In Figure~\ref{fig:HDF_B_local} we put our LF, SLY97 Schechter fit, a
nd Schechter functions reported by 
Metcalfe et al. (1998) and Ellis et al. (1996) (Autofib Redshift Survey) 
The dotted line depicts SLY97 LF, dot-dashed line represents the 
Metcalfe et al. (1998) $B$-band LF, and long dashed line is the Autofib
LF at $z < 0.1$.
Our LF and that of SLY97 agree with higher-normalization LF reported by
Autofib Survey, but are significantly higher than that of 
Metcalfe et al.~(1998)
Autofib LF is also consistent with the LF of ESP Survey (Zucca et al. 1997), 
while Metcalfe et al.'s LF is consistent with EEP LF and Stromlo--APM LF
(Loveday et al.~1992).
But since the solid angle covered by HDF is extremely small and thus the 
normalization can be strongly affected by cosmic variance, we should not
go into further discussion.

We note that the SLY97 $M_*$ value is significantly higher than those of 
other surveys.
This is because the exponential decline at bright end is not observed in 
the HDF LF at $0 < z < 0.5$, and a bump exists at $M_B \sim -20$ mag.
We should also mention that the error bar of the $M_*$ of SLY97 is very
large (1.6 mag).
Considering the large error bars and the uncertainty of the photometric
redshift, we conclude that the bright $M_*$ is not a real feature.

At this lowest redshift, the rise of the faint end is prominent.
The problem of the faint-end slope of galaxy LF has long been a matter of 
debate, and we do not have a widely accepted consensus yet.
As we already pointed out in the above, even in $I$-band we find a 
steepening of the faint end.
If this steep faint end is the artifact of the clustering, the LF derived
from $1/V_{\rm max}$ and those derived from other estimators should have 
been different (Section~\ref{subsec:mock}).
But in fact, they are consistent with each other.
Thus we conclude that, at least in the HDF, the faintest end of the LF has
a steep slope in the Local Universe.

\subsubsection{Evolution at $B$-band: $0.5 < z < 1.0$}

It seems that the brighter galaxies are more numerous than
the local value at $0.5 < z < 1.0$.
Here we should remember the fact that 
the ``fuzzy'' redshift determination is known to affect the shape 
estimation (Liu et al. 1998).
Liu et al. (1998) showed by numerical
experiments that the faint-end slope is underestimated and
$M_*$ is overestimated by the photometric redshift blurring.
The uncertainty of the photometric redshift is rather independent of
the object redshift, so the effect will be severer at the low-$z$, 
and the $M_*$ can be overestimated.
Thus the increase of the bright galaxies is partially due to this effect.
But we can discuss the trend of the LF evolution by comparison of the LF 
derived from photometric redshifts consistently (Liu et al. 1998).

\subsubsection{Evolution at $B$-band: $1.0 < z < 2.0$}

The LFs of the redshift range $1.0 < z < 1.5$ and $1.5 < z < 2.0$ are the 
most reliable ones among the LFs in Figure~\ref{fig:HDFLF_B}, since
the sample size is twice larger than those of the other redshift ranges, 
and in addition, the photometric redshift error becomes worse again at 
$z > 2$.
SLY97 suggested the steepening of the faint-end slope at this redshift.
Our LF of $1.5 < z < 2.0$ presents a similar feature, though the slope 
becomes flatter at the faintest regime.
The deformation of the LF from $z \sim 0$ to $z \sim 2$ supports that 
the steepening of the fainter side of the LF, which is confirmed at $z < 1$,
is continued up to $z \sim 2$.
We do not find a significant shift of $M_*$ at this redshift range.

\subsubsection{Evolution at $B$-band: $2.0 < z$}

The normalization of the furthest redshift LFs settles down 
to the local value, while we also find a brightening of $M_*$ at $z > 3.0$.
We must be careful that in such high redshift, cosmological 
surface brightness dimming is quite severe, and selection effect becomes 
significant (Ferguson 1998; Weedman et al. 1998).
Other kinds of selection effects are discussed in Pascarelle et al. (1998).
Therefore, there can exist more numerous galaxies than estimated.
Further discussions require delicate treatment of such effects.

\subsubsection{Luminosity density evolution}

In order to explore the cosmic star formation history, we derived the 
luminosity density at $B$- and $I$-band based on our LFs.
We fit the Schechter function and extrapolate the faint end below the 
detection limit.
As we mentioned above, the Schechter parameters are poor indicators of the
galaxy evolution, but the integrated luminosity density 
$\rho_{\rm L}$ is regarded as an indicator of the evolution of galaxies, 
because in case the Schechter parameters are significantly affected 
by the fluctuations, $\rho_{\rm L}$ is robust against the effect.
We showed the derived $\rho_{\rm L}$ in Figure \ref{fig:ldensity}.
The upper panel shows the evolution of the $B$-band luminosity density, 
$\rho_{\rm L}(B)$, and the lower, the $I$-band luminosity density, 
$\rho_{\rm L}(I)$.
Open squares are $\rho_{\rm L}(B)$ derived from CFRS (Lilly et al. 1995), 
open circles, $\rho_{\rm L}(B)$ from Autofib (Ellis et al. 1996), 
open triangle represents the value from Stromlo-APM (Loveday et al. 1992)
and open diamond, ESP value (Zucca et al. 1997).
Crosses are the estimates of SLY97.
In this paper we did not try to correct for the reddening effect of dust.

We see the local diversity of the $\rho_{\rm L}(B)$, corresponding to the
normalization discrepancy in Figure \ref{fig:ldensity}. 
Despite the fact that the local LF is hard to derive from the HDF data,
our low-$z$ value is consistent with other previous results.
Added to this, our $\rho_{\rm L}(B)$ at $0.5 < z < 1.0$ significantly 
suffers from the redshift blurring effect, but it is also consistent 
with CFRS highest redshift point within the errors.
As a whole, our result is consistent with that of SLY97, except for 
$2.0 < z < 3.0$.
In this redshift range, $\rho_{\rm L} (B)$ of SLY97 is several times 
larger than our estimate.
This difference may be because SLY97 obtained a steeper $\alpha$ and 
brighter $M_*$ than ours.
We find a flatter LF slope, and the estimates fainter than 
$-20$ mag are not reliable in our result since the fluctuation is 
horribly large at this redshift.
If we choose steeper slope, our $\rho_{\rm L}$ will be higher.
We have to wait for larger datasets to address this problem.
At very high-$z$, we derived a moderately high $\rho_{\rm L}$, 
implying significant numbers of stars have already formed at such 
a high redshift.
The evolution of $\rho_{\rm L}(I)$ appears to be flat.
At the longer wavelength, the observed light is dominated by the
contribution from lower-mass stars, and the temporal change of the SFR 
is less prominent.
We, in addition, should note that the $I$-band results are subject to 
larger $K$-correction extrapolation uncertainties, compared with
the $B$-band results.

At last, we must notice that the above discussions do not account for
the fact that the sample is selected at $I$-band, and 
the selection criterion is different for each redshift range.
At $z < 0.5$, the sample is safely regarded as $I$-selected, while
at $z > 3.0$, they are in fact rest UV-selected.
Thus the ideal discussion on the evolution of the galaxy LF 
should be based on the suitably designed survey as performed 
by Cowie et al. (1999).
We will consider this point, and make more sophisticated discussions 
elsewhere (Takeuchi 2000, in preparation).

\section{SUMMARY AND CONCLUSION}\label{sec:conclusion}

The estimation of the LF from observational data is not a trivial task,
because of the flux-limited nature of the astronomical data.
We focused on the following four estimators: 
1) Schmidt--Eales ($1/V_{\rm max}$) method
, 2) Efstathiou--Ellis--Peterson (EEP) method, 3) Cho{\l}oniewski method, 
and 4) Lynden-Bell--Cho{\l}oniewski--Caditz--Petrosian (LCCP) method.
We improved some of the estimators for studying the very distant universe, 
and examined their performances for much wider class of functional forms 
by Monte Carlo simulation.
We tested these four estimators by the numerical experiments with 
mock catalogs.
We also used the mock 2dF catalogs prepared by Cole et al. (1998).
Then we applied these estimators to the HDF photometric redshift catalog
of Fern\'{a}ndez-Soto et al. (1999).
Our conclusions are as follows:
\begin{enumerate}
\item If the sample is spatially homogeneous, all estimators give 
consistent results with each other, and we did not find any bias for 
any LF shapes.
Thus, when we have a sufficiently large galaxy sample, we can use any of 
the estimators examined in this paper.
Even when the sample size is smaller, the mean values remain unbiased, 
though the standard deviations become larger.

\item Large fluctuation appears at the faint end of the LF, because the 
amount of available data is small.
Therefore, the flatter the LF slope is, the larger the fluctuations become.
When the sample size is small, fluctuations in the Cho{\l}oniewski method 
become seriously large due to shot noise, and thus we recommend this method 
for the analysis of large samples.

\item When a large cluster or void exists, $1/V_{\rm max}$ estimator is 
severely affected in its LF shape estimation.
The other three estimators are not affected by a cluster or void at all.
They gave consistent results with each other, and the estimates showed
perfect agreement with the input LFs.

\item We examined the calculation time of each method.
Because of its algorithmic simplicity, Cho{\l}oniewski method is the fastest
among the four methods.
The EEP method needs more iterations in the procedure than others do, 
and longer calculation time.
The $1/V_{\rm max}$ method calculates the maximum volume $V_{\rm max}$ for 
each galaxy, and also needs significant calculation time.
The LCCP method requires a large stack for data sorting procedure, which 
is a requirement of this method.
Thus we stress that the Cho{\l}oniewski method is the most economic from 
the standpoint of practical computing.

\item We examined more realistic large mock samples, specifically 
mock 2dF catalogs prepared by Cole et al. (1998).
We found that the redshift distortion does not affect the LF estimates.
When we treat such a large catalog, the advantage of the Cho{\l}oniewski
method is extremely significant in terms of the computation time.

\item We derived the $I$- and $B$-band luminosity function of the HDF.
The four different LF estimators gave consistent results for the HDF sample.
We found the overall brightening of the LF. 
It seems that the faint-end steepens toward $z = 2 - 3$, 
and settles down to the local value at $z \sim 3$.
We note that the ``fuzzy'' redshift determination is known to affect the 
shape estimation (Liu et al. 1998).

\item We found a rather mild evolution of the LF.
Despite the fact that the local LF is hard to derive from the HDF data,
our low-$z$ value is consistent with other previous results.
Our $\rho_{\rm L}(B)$ at $0.5 < z < 1.0$ is also consistent 
with CFRS highest redshift point within the errors.
As a whole, our result is roughly consistent with that of SLY97, but lower at 
$2.0 < z < 3.0$.
At very high-$z$, we derived a moderately high $\rho_{\rm L}$, 
implying that a significant numbers of stars have already formed at such 
a high redshift.
We found that the evolution of $\rho_{\rm L}(I)$ is flat.
\end{enumerate}

\acknowledgments
First we would like to thank the anonymous referee for his useful 
suggestions and comments, which improved our paper very much in its 
clarity and English presentation.
We offer our gratitude to Hiroyuki Hirashita and Fumiko Eizawa
who gave a lot of useful suggestions.
We also thank Kouji Ohta, Kouichiro Nakanishi, Toru Yamada,
Takashi Ichikawa, Kazuhiro Shimasaku for their fruitful discussions 
and useful comments.
Mamoru Sait\={o}, Hiroki Kurokawa and Yasushi Suto are thanked for their 
continuous encouragements.
This work owes a great debt to the photometric redshift catalog prepared
by Fern\'{a}ndez-Soto et al. 
TTT and KY acknowledge the Research Fellowships
of the Japan Society for the Promotion of Science for Young
Scientists.
We carried out the numerical computations and extensively used the databases
at the Astronomical Data Analysis Center of the National
Astronomical Observatory, Japan, 
which is an inter-university research institute of astronomy operated by 
Ministry of Education, Science, Culture, and Sports. 
\appendix

\section{AKAIKE'S INFORMATION CRITERION}\label{sec:aic}

In this appendix, we make an informal introduction of 
Akaike's theory.
The meaning of the maximum likelihood method is clearly understood by 
using the concepts of information theory.
Since the middle of 1970's, vast advances have been made in the field of 
the statistical inference by the discovery of Akaike's Information Criterion 
(AIC:  Akaike 1974).
The AIC is closely related to the information entropy, especially to the 
`relative entropy' of two probability distributions.
The relative entropy has a property just like a distance in differential 
geometry, i.e. it is a distance between the two probability distributions.
Using AIC enables us to compare the goodness of a certain model with that of
another type directly.
For this fascinating property, AIC is applied to various fields of studies.
The AIC is expressed as 
\begin{eqnarray}\label{eq:aic}
{\rm AIC} = -2(\ln {\cal L}(\hat{\theta})-K)\; ,\nonumber
\end{eqnarray}
where $\cal L$ is a likelihood function, $\hat{\theta}$ is a set of maximum 
likelihood estimators, and $K$ is the number of free parameters of the 
assumed model.
The ``most preferred'' model is the one which minimizes the AIC.

Here we present the problem of polynomial regression model selection
by using AIC as an example.
As we mentioned in Section~\ref{sec:hdf}, we adopted this procedure 
to determine the order of $K$-correction as a function of redshift.
Given a set of $n$ pairs of observations 
$(x_1, y_1), \; \cdots, \;(x_n, y_n)$, we fit the $m$-th order polynomial
model 
\begin{eqnarray}
  y_i = \sum_{\ell = 0}^m a_\ell x_i^\ell + \varepsilon_i \;,
\end{eqnarray}
where $\varepsilon_i$ is an independent random variable 
which follows the normal distribution with mean $0$ and dispersion $\sigma^2$.
The variables $x_i$ and $y_i$ are called the explanatory variable and the 
objective variable, respectively.
This model is a conditional distribution of which the distribution 
of the objective variable $y$ is a normal distribution $f(y_i)$ with the mean 
$a_0 + a_1x_i + \cdots + a_mx_i^m$ and the variance $\sigma^2$, i.e.
\begin{eqnarray}
  f(y_i | a_0, \cdots , a_m, \sigma^2) = \cfrac{1}{\sqrt{2\pi \sigma^2}}
  \exp \left\{ \frac{1}{2\sigma^2} \left( 
      y_i - \sum_{\ell = 0}^m a_\ell x_i^\ell \right)^2 \right\} \;.
\end{eqnarray}
Therefore, when a set of data is $(x_1, y_1), \; \cdots, \;(x_n, y_n)$, 
the likelihood is given by 
\begin{eqnarray}
  {\cal L} (y_1, \cdots , y_n | a_0, \cdots , a_m, \sigma^2) 
  &=& \prod_{i = 1}^{n} f(y_i | a_0, \cdots , a_m, \sigma^2) \nonumber \\
  &=& \left( \cfrac{1}{\sqrt{2\pi \sigma^2}} \right)^\frac{n}{2}
  \prod_{i = 1}^{n} \exp \left\{ \frac{1}{2\sigma^2} \left( 
      y_i - \sum_{\ell = 0}^m a_\ell x_i^\ell \right)^2 \right\} \;.
\end{eqnarray}
The log likelihood is then expressed as
\begin{eqnarray}\label{eq:polylikelihood}
  \ln {\cal L} (y | a_0, \cdots , a_m, \sigma^2) 
  &=& -\frac{n}{2} \ln 2\pi -\frac{n}{2} \ln \sigma^2 
  - \frac{1}{2\sigma^2} \sum_{i = 1}^{n} 
  \left(y_i - \sum_{\ell = 0}^m a_\ell x_i^\ell \right)^2 \;.
\end{eqnarray}
The log likelihood eq.~(\ref{eq:polylikelihood}) is maximized with respect to
$a_0, \cdots, a_m$ when
\begin{eqnarray}
  S \equiv \sum_{i = 1}^{n} 
  \left(y_i - \sum_{\ell = 0}^m a_\ell x_i^\ell \right)^2 \;.
\end{eqnarray}
is minimized.
Thus, in the case of polynomial model fitting, the maximum likelihood
procedure is equivalent to the least square method.
The necessary conditions that $a_0, \cdots, a_m$ maximize $S$ are 
the normal equations of the least square,
\begin{eqnarray}
  \pa{S}{a_0} &=& -2 \sum_{i = 1}^{n} 
  \left(y_i - \sum_{\ell = 0}^m a_\ell x_i^\ell \right) = 0 \nonumber \\
  \pa{S}{a_1} &=& -2 \sum_{i = 1}^{n} x_i
  \left(y_i - \sum_{\ell = 0}^m a_\ell x_i^\ell \right) = 0 \nonumber \\
  &\vdots& \nonumber \\
  \pa{S}{a_m} &=& -2 \sum_{i = 1}^{n} x_i^m
  \left(y_i - \sum_{\ell = 0}^m a_\ell x_i^\ell \right) = 0 \; ,
\end{eqnarray}
and the maximum likelihood estimates $\hat{a_0}, \cdots, \hat{a_m}$ are 
obtained by solving these linear equations.
Besides, the necessary condition that $\sigma^2$ maximizes 
eq.~(\ref{eq:polylikelihood}) is
\begin{eqnarray}
  \pa{\ln {\cal L}}{\;\sigma^2} 
  = -\frac{n}{2\sigma^2} + \frac{1}{2(\sigma^2)^2} \sum_{i = 1}^{n} 
  \left(y_i - \sum_{\ell = 0}^m \hat{a_\ell} x_i^\ell \right)^2 = 0\;.
\end{eqnarray}
The maximum likelihood estimate of the residual variance $\sigma^2$ is 
\begin{eqnarray}\label{eq:mresidual}
  \hat{\sigma^2} = -\frac{1}{n} \sum_{i = 1}^{n} 
  \left(y_i - \sum_{\ell = 0}^m \hat{a_\ell} x_i^\ell \right)^2
  = -\frac{1}{n} \left( \sum_{i = 1}^{n} y_i^2 - 
    \sum_{\ell = 0}^{m} \hat{a_\ell} \sum_{i = 1}^{n}x_i^\ell y_i \right)\;.
\end{eqnarray}
Hereafter we denote the residual variance $\sigma^2$ for a model with 
$m$-th order as $\sigma^2(m)$.
Then, from eqs.~(\ref{eq:polylikelihood}) and (\ref{eq:mresidual}), 
the maximum log likelihood becomes
\begin{eqnarray}\label{eq:polymaxl}
  \ln {\cal L} (y | \hat{a_0}, \cdots , \hat{a_m}, \hat{\sigma^2}) 
  = -\frac{n}{2} \ln 2\pi- \frac{n}{2} \ln \hat{\sigma^2}(m) - \frac{n}{2}\;.
\end{eqnarray}
The $m$-th order polynomial model has $m+2$ parameters 
($a_0, \cdots, a_m, \sigma^2(m)$).
Substituting $K = m+2$ and eq.~(\ref{eq:polymaxl}) into eq.~(\ref{eq:aic})
gives the AIC of the $m$-th order model,
\begin{eqnarray}
  {\rm AIC} (m) = n (\ln 2\pi + 1)+ n \ln \hat{\sigma^2}(m) + 2(m+2)\;.
\end{eqnarray}

The result of our polynomial fitting to the $K$-correction is presented
in Figure~\ref{fig:polyfit}.
We also summarize the AIC value for each polynomial order in 
Table~\ref{table:aic}.
The AIC values of elliptical, Sbc, and Irr in table~\ref{table:aic} 
really took their minima in the case that the fitting polynomials were
those with 5th order, and only Scd data preferred the 6th order.
Putting all accounts together, we chose 5th order polynomial model.

\newpage

\begin{center}
{\large \bf Figure Captions}\footnote[1]{\large All figures listed below are
available from ftp://ftp.kusastro.kyoto-u.ac.jp/pub/kohji/lf/~.}
\end{center}

\figurenum{1}
\figcaption{
Schematic description of the Cho{\l}oniewski's method.
}\label{fig:choloniewski}

\figurenum{2}
\figcaption{
Schematic description of the Lynden-Bell--Cho{\l}oniewski--Caditz--Petrosian 
method.
}\label{fig:lccp}

\figurenum{3}
\figcaption{
Simulated functional forms for the luminosity functions.
The density scale is arbitrary because it depends on the adopted sample size,
and we set the scale for the case of size $\sim 1000$.
}\label{fig:lfforms}

\figurenum{4}
\figcaption{
An example of the spatial configuration of galaxies of a cluster sample.
}\label{fig:clustersample}

\figurenum{5a}
\figcaption{
Estimates of the luminosity function (LF) for mock samples with size 
$\sim 1000$.
The input luminosity function is with the functional form A.
This is the LF of the spatially homogeneous sample.
}\label{fig:mock_uu_1000}

\figurenum{5b}
\figcaption{
The same as Figure~\ref{fig:mock_uu_1000}, except that this is the LF 
estimated from the cluster sample.
}\label{fig:mock_uc_1000}

\figurenum{5c}
\figcaption{
The same as Figure~\ref{fig:mock_uu_1000}, except that this is the LF 
estimated from the void sample.
}\label{fig:mock_uv_1000}

\figurenum{6a}
\figcaption{
The same as Figure~\ref{fig:mock_uu_1000} except that the LF shape is B.
}\label{fig:mock_fu_1000}

\figurenum{6b}
\figcaption{
The same as Figure~\ref{fig:mock_uc_1000} except that the LF shape is B.
}\label{fig:mock_fc_1000}

\figurenum{6c}
\figcaption{
The same as Figure~\ref{fig:mock_uv_1000} except that the LF shape is B.
}\label{fig:mock_fv_1000}

\figurenum{7a}
\figcaption{
The same as Figure~\ref{fig:mock_uu_1000} except that the LF shape is C.
}

\figurenum{7b}
\figcaption{
The same as Figure~\ref{fig:mock_uc_1000} except that the LF shape is C.
}

\figurenum{7c}
\figcaption{
The same as Figure~\ref{fig:mock_uv_1000} except that the LF shape is C.
}

\figurenum{8a}
\figcaption{
The same as Figure~\ref{fig:mock_uu_1000} except that the LF shape is D.
}

\figurenum{8b}
\figcaption{
The same as Figure~\ref{fig:mock_uc_1000} except that the LF shape is D.
}

\figurenum{8c}
\figcaption{
The same as Figure~\ref{fig:mock_uv_1000} except that the LF shape is D.
}

\figurenum{9a}
\figcaption{
The same as Figure~\ref{fig:mock_uu_1000} except that the LF shape is E.
}

\figurenum{9b}
\figcaption{
The same as Figure~\ref{fig:mock_uc_1000} except that the LF shape is E.
}

\figurenum{9c}
\figcaption{
The same as Figure~\ref{fig:mock_uv_1000} except that the LF shape is E.
}

\figurenum{10a}
\figcaption{
The same as Figure~\ref{fig:mock_uu_1000}  except that the LF shape is F.
}

\figurenum{10b}
\figcaption{
The same as Figure~\ref{fig:mock_uc_1000}  except that the LF shape is F.
}

\figurenum{10c}
\figcaption{
The same as Figure~\ref{fig:mock_uv_1000}  except that the LF shape is F.
}

\figurenum{11a}
\figcaption{
Estimates of the luminosity function (LF) for mock samples with size 100.
The input luminosity function is with the functional form A.
This is the LF of the spatially homogeneous sample.
}\label{fig:mock_uu_100}

\figurenum{11b}
\figcaption{
The same as Figure~\ref{fig:mock_uu_100}, but this is the LF from the
cluster sample.
}\label{fig:mock_uc_100}

\figurenum{11c}
\figcaption{
The same as Figure~\ref{fig:mock_uu_100}, but this is the LF from the
void sample.
}\label{fig:mock_uv_100}

\figurenum{12a}
\figcaption{
The same as Figure~\ref{fig:mock_uu_100} except that the LF shape is B.
}\label{fig:mock_fu_100}

\figurenum{12b}
\figcaption{
The same as Figure~\ref{fig:mock_uc_100} except that the LF shape is B.
}\label{fig:mock_fc_100}

\figurenum{12c}
\figcaption{
The same as Figure~\ref{fig:mock_uv_100} except that the LF shape is B.
}\label{fig:mock_fv_100}

\figurenum{13a}
\figcaption{
The same as Figure~\ref{fig:mock_uu_100} except that the LF shape is C.
}

\figurenum{13b}
\figcaption{
The same as Figure~\ref{fig:mock_uc_100} except that the LF shape is C.
}

\figurenum{13c}
\figcaption{
The same as Figure~\ref{fig:mock_uv_100} except that the LF shape is C.
}

\figurenum{14a}
\figcaption{
The same as Figure~\ref{fig:mock_uu_100} except that the LF shape is D.
}

\figurenum{14b}
\figcaption{
The same as Figure~\ref{fig:mock_uc_100} except that the LF shape is D.
}

\figurenum{14c}
\figcaption{
The same as Figure~\ref{fig:mock_uv_100} except that the LF shape is D.
}

\figurenum{15a}
\figcaption{
The same as Figure~\ref{fig:mock_uu_100} except that the LF shape is E.
}

\figurenum{15b}
\figcaption{
The same as Figure~\ref{fig:mock_uc_100} except that the LF shape is E.
}

\figurenum{15c}
\figcaption{
The same as Figure~\ref{fig:mock_uv_100} except that the LF shape is E.
}

\figurenum{16a}
\figcaption{
The same as Figure~\ref{fig:mock_uu_100}  except that the LF shape is F.
}

\figurenum{16b}
\figcaption{
The same as Figure~\ref{fig:mock_uc_100}  except that the LF shape is F.
}

\figurenum{16c}
\figcaption{
The same as Figure~\ref{fig:mock_uv_100}  except that the LF shape is F.
}

\figurenum{17}
\figcaption{
The input and estimated LFs of the mock 2dF catalog.
The cosmological parameters are $\Omega_0 = 1.0$, 
$\lambda_0 = 0$, $\Gamma =1$, and $\sigma_8 = 0.55$.
}\label{fig:2dFLF_EdS}

\figurenum{18}
\figcaption{
The same as Figure~\ref{fig:2dFLF_EdS}, 
except that the cosmological parameters are 
$\Omega_0 = 0.3$, $\lambda_0 = 0.7$, $\Gamma = 0.25$, and $\sigma_8 = 1.13$.
}\label{fig:2dFLF_L3S}

\figurenum{19}
\figcaption{
The same as Figure~\ref{fig:2dFLF_EdS}, 
except that the cosmological parameters are 
$\Omega_0 = 0.3$, $\lambda_0 = 0$, $\Gamma = 0.25$, and $\sigma_8 = 1.13$.
}\label{fig:2dFLF_O3S}

\figurenum{20}
\figcaption{
The $I$-band LF derived from the HDF photometric redshift 
catalog prepared by Fern\'{a}ndez-Soto et al.~(1999).
The dotted line represents the local $I$-band LF 
obtained by Metcalfe et al.~(1998).
}\label{fig:HDFLF_I}

\figurenum{21}
\figcaption{
The same as Figure~\ref{fig:HDFLF_I}, except that this is the $B$-band LF.
The dotted lines depict the $B$-band LFs derived by 
Sawicki et al.~(1997).
}\label{fig:HDFLF_B}

\figurenum{22}
\figcaption{
The $B$-band LF of the HDF at the redshift range $0 < z < 0.5$.
The dotted line depicts the Sawicki et al.~(1997) LF,  
dot-dashed line represents the Metcalfe et al.~(1998)
$B$-band LF, and long dashed line is the Autofib LF at $z < 0.1$.
Our LF and that of Sawicki et al.~(1997) agree with higher-normalization 
LF reported by Autofib Survey, but are significantly higher than that of 
Metcalfe et al. (1998).
}\label{fig:HDF_B_local}

\figurenum{23}
\figcaption{
The evolution of the luminosity density, $\rho_{\rm L}$.
Upper panel: the $B$-band luminosity density.
Lower panel: the $I$-band luminosity density.
Open squares are $\rho_{\rm L}(B)$ derived from CFRS (Lilly et al. 1995), 
open circles, $\rho_{\rm L}(B)$ from Autofib (Ellis et al. 1996), 
open triangle represents the value from Stromlo-APM (Loveday et al. 1992)
and open diamond, ESP value (Zucca et al. 1997).
Crosses are the estimates of Sawicki et al. (1997).
In this paper we did not try to correct for the reddening effect of dust.
}\label{fig:ldensity}

\figurenum{24}
\figcaption{
The result of the polynomial fitting to the $K$-correction of galaxies
constructed from the data prepared by Kinney et al.~(1996).
Symbols represent the representative SEDs of galaxies 
(open squares: elliptical, open triangles: Sbc, diamonds: Scd, and crosses:
Irr).
Panels (a) -- (f) correspond to the fitting order 1st -- 6th, 
respectively.
}\label{fig:polyfit}

\pagebreak

\tablenum{1}
\begin{table}
\label{table:aic}
\begin{center}
Table 1: The AIC of the polynomial fitting model for $K$-correction

\begin{tabular}{lrrrrrr}
\tableline \tableline
Type & \multicolumn{6}{c}{Polynomial fitting order}\\
~ & 1st & 2nd & 3rd & 4th & 5th & 6th \\ \tableline
Elliptical & 96.4724  &   41.6174 &   42.9716 &   19.0966 &  $-$6.93137 &   2.50965 \\
Sbc & 63.4433  &  $-$9.27677 &  $-$7.30190 &  $-$26.7273 &  $-$45.7942 &  $-$32.1221 \\
Scd & 5.04120  &   4.66447 & $-$0.536911 &  $-$12.5160 &  $-$11.0097 &  $-$24.8293 \\
Irr & $-$7.86433 &  $-$31.3419 &  $-$31.0330 &  $-$49.8901 &  $-$65.2359 &  $-$57.3888 \\ \tableline
\end{tabular}
\end{center}
\end{table}

\end{document}